 \documentclass[aps,pra,showpacs,preprintnumbers,amsmath,amssymb]{revtex4}
 \usepackage{graphicx}
 \usepackage{dcolumn}
 \usepackage{bbm}
 \usepackage{color}

\begin{document}

 \bibliographystyle{apsrev}

 \preprint{APS/123-QED}
 \title{Vibrations and damping of the eigenmodes of viscoelastic nanospheres with thermal conductivity}

 \author{M. Wenin}
 \email{markus.wenin@cphysics.com}

 \affiliation{%
\\ CPE Computational Physics and Engineering, Weingartnerstrasse 28, 39011 Lana, BZ, Italy
 }%

 \author{A. Windisch}
 \email{windisch@physics.wustl.edu}

 \affiliation{%
\\Physics Department, Washington University in St. Louis, MO, 63130, USA
 }%
 \affiliation{%
\\DIGITAL Institute for Digital Technologies, JOANNEUM RESEARCH Forschungsgesellschaft mbH, Graz, Austria
 }%
 \date{\today}

\begin{abstract}
In this paper, we derive a series of exact analytical closed expressions to calculate the complex eigenfrequencies and the displacement for the corresponding eigenmodes of a viscoelastic (nano-) sphere in the presence of linear damping. Where possible, we provide closed expressions for damping rates, including the contributions from viscosity, as well as thermal conductivity and solutions of the heat equation. We assume an isolated system, such that no energy/heat transfer to the environment is allowed. We find monotonic behavior of the damping as a function of frequency for breathing and torsional modes, however, for spheroidal modes we find non-monotonicity. Furthermore, we analytically analyze the thermodynamic limit for all mode types. We also investigate the frequency shift and find expected behavior, i.e. a reduced eigenfrequency with damping than without damping for breathing and torsional modes. For spheroidal modes, however, we find non--monotonic  shifts, corresponding to the damping. For some eigenfrequencies we find anomalous frequency shifts.    
\end{abstract}

 \maketitle
\section{Introduction}
In this paper we investigate the damping and temperature distribution of a nano sphere using classical continuum physics \cite{Landau7}. Analytical studies of homogeneous spheres  have been conducted in different contexts, see e.g. in geophysics \cite{Ben-Menahem1981,Usami}, or the physics of older gravitational wave detectors \cite{COCCIA1996263,Siegel2010ExcitationON}. Here we focus on the application to metallic nano spheres with non-negligible thermal conductivity, taking into account the natural limits of applicability of classical continuum mechanics \cite{Erbi,Ducottet_2023}.\\ In order to deal with a well defined simplified problem, we restrict ourselves to the case of a free, isotropic, thermally isolated sphere. Other boundary conditions, as well as spheres with anisotropic material properties, or embedded in fluids, are investigated in e.~g. \cite{Anderson,ElBaroudi,Schafbuch,Fatti,Buchanan,Tian}. 
The eigenfrequencies and -modes for isotropic and homogeneous elastic spheres are completely known, see e.~g. \cite{Lamb,Eringen,Dickinson,Rue}. We re-derive these results with the help of the computer algebra system Mathematica \cite{mathematica}, where we include damping by a linear dissipation tensor, and obtain the frequencies and the displacement fields. In a subsequent step, we derive the damping coefficients using the dissipation function and the contribution from thermal conductivity. We consider three different approaches: the standard approach of complex eigenfrequencies, a Taylor expansion method taking into account viscosity only, and the analytical or numerical evaluation of integral expressions to obtain the viscosity and the thermal part of the damping.
The heat equation allows for the study of the temperature distribution inside the sphere when vibrating in an eigenmode, in which the adiabatic compression/expansion leads to local heating/cooling of the material. Since we neglect temperature gradients in the stress tensor, we are able to solve the elastic problem and the thermal problem independently (for a theoretical study of the coupled system see \cite{Sharma,OpenAcoustic}).\\
Our paper provides a complete collection of analytic results for the damping of eigenmodes for all symmetry types: breathing modes, spheroidal modes, and torsional modes.\\
These systems are also studied experimentally by Laser induced excitation and subsequent measurement of vibrations by Raman/Brillouin scattering \cite{PhysRevB.69.113402,PhysRevB.72.205433,JuanHernndezRosas2003,Galstyan,Hodak,10.1063/5.0146969}. 
Our work is inspired partially by these experiments, together with the research question, whether finite size effects play a significant role in damping effects.  
The paper is organized as follows: First, we present the basic theory, definitions, and governing equations important for this study. Next we consider the viscoelastic sphere and give the main results for the displacement field, the damping coefficients, and the temperature field. We provide the reader with a series of exact, closed analytical results. Furthermore we discuss also the frequency shifts, in particular for the spheroidal modes, based on the exact solution of the viscoelastic problem including the boundary condition. The obtained results (complex frequencies) are quite interesting for either pure mathematical considerations regarding the Stokes equation, or physical considerations regarding the relationship between viscosity and frequency \cite{Buckingham,Destrade}.
Detailed proofs can be found in our accompanying Mathematica notebook \textit{DampingViscoelasticSphere.nb}.\\ A section is devoted to numerical computations to illustrate and support the analytical results, where we consider a Ag nano sphere. The two appendices contain further technical details on the derivation of the solution, as well as definitions of auxiliary quantities used throughout the paper.     
\section{Theory}\label{sect1}
\subsection{Basic definitions and equations of motion}
We consider a viscoelastic sphere with radius $R$. Let $\mathbf{\zeta}(\mathbf{x},t)$ be the displacement field, where we express the spatial dependence in spherical coordinates $\mathbf{x} = \{r,\theta,\phi\}$. The respective basis vectors are denoted as $\{\mathbf{e}_r,\mathbf{e}_\theta,\mathbf{e}_\phi\}$.
The elastic stress tensor contains the two adiabatic elastic moduli $\lambda_1$ and $\lambda_2$, and reads,
\begin{equation}\label{sigma}
\sigma_{\mathrm{el}} = \lambda_1 \left[\nabla\otimes\mathbf{\zeta} + (\nabla\otimes\mathbf{\zeta})^T\right] + \lambda_2 (\nabla\cdot\mathbf{\zeta}) \cdot \mathbbm{1}~.
\end{equation}
Here and in the following, we use the symbol $\mathbbm{1}$ to denote the $3\times 3$ unit matrix. It is useful to introduce the strain tensor,
\begin{equation}\label{Z}
\varepsilon = \frac{1}{2}\left[\nabla\otimes\mathbf{\zeta} + (\nabla\otimes\mathbf{\zeta})^T\right]~.
\end{equation}
Dissipation is described by the viscosity tensor,
\begin{equation}\label{sigmas}
\sigma_{\mathrm{diss}} = 2\eta_1 \dot{\varepsilon} + \eta_2 (\nabla\cdot\dot{\mathbf{\zeta}}) \cdot \mathbbm{1}~,
\end{equation}
which contains the phenomenological shear viscosity coefficient $\eta_1=\eta_S\geq 0$, and the second viscosity coefficient $\eta_2$, which is related to the volume viscosity, $\zeta_V = \eta_2+2/3\eta_1\geq 0$, see \cite{Schmutzer,Landau7}. 
The equation of motion for the continuum, and with constant mass density $\rho$, is given by
\begin{equation}\label{eqzeta}
\rho \frac{\partial^2 \mathbf{\zeta}}{\partial t^2}=\nabla\cdot\sigma~,\hspace{5mm}\sigma=\sigma_{\mathrm{el}}+\sigma_{\mathrm{diss}}~.
\end{equation}
We use the boundary condition of a tensionless state at $r=R$, $\sigma\cdot\mathbf{e}_r = \mathbf{0}$. 
In our investigations we furthermore consider the equation of motion for the temperature field $\delta T(\mathbf{x},t)$, measured relative to a reference temperature $T_0$. We use the inhomogeneous heat equation to determine $\delta T(\mathbf{x},t)$ inside the sphere. The time--dependent displacement field acts as a reversible heat source. Thermal conductivity and viscosity lead to irreversible processes. To calculate $\delta T(\mathbf{x},t)$ we only take the compression/expansion of the material into account, and not viscosity. Thus, with $\alpha$ being the thermal diffusivity, we have
\begin{equation}\label{basicT}
\frac{\partial
\delta T(\mathbf{x},t)}{\partial
t} - \alpha \triangle \delta T(\mathbf{x},t) = T_v \nabla\cdot\dot{\mathbf{\zeta}}~,
\end{equation}
augmented by an adiabatic boundary condition at the surface $r=R$, $(\partial\delta T/\partial r)_{r=R}=0$.
In Eq.~\eqref{basicT} we use the constant
\begin{equation}
T_v =-\frac{\alpha_T T_0 K}{c_P\rho}~,
\end{equation}
where $\alpha_T$ is the thermal volume expansion coefficient, $K=\frac{2}{3}\lambda_1+\lambda_2$ the modulus of compressibility, and $c_P$ the heat capacity. We also introduce the thermal conductivity $\mathrm{k}=\rho\alpha c_P$.
\subsection{Equations for energy and dissipation}
We need the expressions for the free energy and the dissipation function for the continuum. For the free energy density (elastic potential) we have
\begin{equation}\label{epot}
f_{\mathrm{el}}(\mathbf{x},t) =\frac{1}{2}\mathrm{Tr}\Big\{\sigma_{\mathrm{el}}\cdot \varepsilon \Big\}~,
\end{equation}
where here and in the following `Tr' denotes the trace of the matrix. For completeness we also give the formula for the kinetic energy density,
\begin{equation}\label{ekin}
e_{\mathrm{kin}}(\mathbf{x},t) =\frac{\rho~\dot{\mathbf{\zeta}}^2}{2}~.
\end{equation}
The dissipation function takes into account the mechanical energy loss due to viscosity,
\begin{equation}\label{dissip}
q(\mathbf{x},t)= \mathrm{Tr}\Big\{ \sigma_{\mathrm{diss}}\cdot\dot{\varepsilon} \Big\}~.
\end{equation}
The total dissipation $\dot{E}_{\mathrm{diss}}\leq 0$ consists of two parts, a thermal part, and a viscosity part: 
\begin{equation}
\dot{E}_{\mathrm{diss}} = \dot{E}_{\mathrm{diss,T}}+\dot{E}_{\mathrm{diss,visc}}=-\frac{\mathrm{k}}{ T_0}\int |\nabla T|^2 d^3\mathbf{x} - \int q d^3\mathbf{x}~.
\end{equation}
To include them both in a single damping constant, we relate the dissipation energy to the total energy of the sphere. 
We integrate the densities Eqs.~(\ref{epot},\ref{ekin}) over the whole sphere and average over one vibration period, which we indicate with angular brackets $\langle...\rangle$. Because of the linear approach in Eq.~\eqref{sigmas} we can neglect dissipation when calculating the corresponding integrals. Furthermore, for this calculation we also put the thermal diffusivity $\alpha=0$ in Eq.~\eqref{basicT}, and the source $q(\mathbf{x},t)=0$, to determine the temperature field inside the sphere, $\delta T = T_v \nabla\cdot\mathbf{\zeta}$. We define the damping constants as 
\begin{eqnarray}\label{gdissint}
\gamma_{\mathrm{diss,visc}}:=\frac{|\langle\dot{E}_{\mathrm{diss,visc}}\rangle|}{2E}~,
\end{eqnarray}
and
\begin{equation}
\gamma_{\mathrm{diss,T}}:=\frac{|\langle\dot{E}_{\mathrm{diss,T}}\rangle|}{2E}~,
\end{equation}
where we used the definition
\begin{equation}
E:= \int \langle f_{\mathrm{el}}+e_{\mathrm{kin}}\rangle d^3 \mathbf{x}~.
\end{equation}
With the time--independent displacement and strain tensor given in Eq.~\eqref{u} below, we obtain the general expressions (the total energy for a dissipation--less system in an eigenstate $\mathbf{u}$ with eigenfrequency $\omega$ is $E=\frac{\rho \omega^2}{2} \int  |\mathbf{u}|^2 d^3\mathbf{x}$)
\begin{equation}\label{gV}
\gamma_{\mathrm{diss,visc}} = \frac{1}{2\rho}\frac{\int\Big[2\eta_1 \mathrm{Tr}\{\tilde{\epsilon}^\dagger\tilde{\epsilon}\}+\eta_2|\nabla\cdot\mathbf{u}|^2\Big]d^3\mathbf{x}}{\int |\mathbf{u}|^2 d^3\mathbf{x}}~,
\end{equation}
and
\begin{equation}\label{gT}
\gamma_{\mathrm{diss,T}} = \frac{\mathrm{k}T_v^2}{2\rho T_0\omega^2}\frac{\int |\nabla(\nabla\cdot \mathbf{u})|^2d^3\mathbf{x}}{\int |\mathbf{u}|^2 d^3\mathbf{x}}~,
\end{equation}
\begin{equation}
\gamma_{\mathrm{diss,tot}}=\gamma_{\mathrm{diss,visc}}+\gamma_{\mathrm{diss,T}}~.
\end{equation}
For the purpose of readability we do not show the eigenfrequency dependence. Note that both $\gamma_{\mathrm{diss,visc}}$ and $\gamma_{\mathrm{diss,T}}$ are strictly positive by construction and valid in principle for any value of $\omega>0$. 
In the following, we provide an alternative derivation of $\gamma_{\mathrm{diss,visc}}$. Let $f(\omega+i\gamma;\eta_1,\eta_2)$ be a function that expresses the eigenfrequencies of a particular vibrational mode (see  Eq.~(\ref{Lambq}), Eq.~\eqref{omegasph} and Eq.~\eqref{omegators} for the actual expressions for breather modes, spheroidal modes, and torsional modes respectively). Conducting a Taylor expansion around the point $\omega=\omega_0,~\gamma=0$ on the real axis, and putting $\eta_1=0,~\eta_2=0$, to first order we obtain
\begin{equation}\label{gValternativ}
\gamma_{\mathrm{diss,visc}}=i\frac{\eta_1\Big(\frac{\partial f}{\partial\eta_1}\Big)_0 + \eta_2\Big(\frac{\partial f}{\partial\eta_2}\Big)_0}{\Big(\frac{\partial f}{\partial\omega}\Big)_0}~,
\end{equation}
where the index zero denotes the expansion point, and $\omega_0$ is one of the real roots of the function $f(\omega_0;0,0)$. Note furthermore, that in contrast to Eq.~\eqref{gV}, Eq.~\eqref{gValternativ} is valid only in the neighborhood around the resonance frequency $\omega_0$. However, in order to not complicate the notation any further, we use the same symbol $\gamma_{\mathrm{diss,visc}}$ as before. To avoid any confusion between these two expressions, we provide specific remarks stating which expression has been used throughout the paper.   
Note, that if we define $\tilde{f}(\omega+i\gamma;\eta_1,\eta_2):=g(\eta_1,\eta_2)f(\omega+i\gamma;\eta_1,\eta_2)$, we have another valid function to determine the eigenfrequencies, because $\tilde{f}(\omega+i\gamma;\eta_1,\eta_2)$ has the same zeros as $f(\omega+i\gamma;\eta_1,\eta_2)$. Thus, Eq.~\eqref{gValternativ} can be used with $\tilde{f}(\omega+i\gamma;\eta_1,\eta_2)$ as well. This is because the additional terms arising from the partial derivatives are zero at the point $(\omega_0,0,0)$. We are stressing this point because different Computer Algebra systems/versions can produce different representations of essentially the same expression.   
\subsection{Equations for the displacement field}
It is common to separate the displacement field into two parts,
\begin{equation}\label{basicutot}
\mathbf{\zeta}(\mathbf{x},t)=\mathbf{\zeta}_l(\mathbf{x},t)+\mathbf{\zeta}_t(\mathbf{x},t)~,
\end{equation}
where $\mathbf{\zeta}_l(\mathbf{x},t)$ is the longitudinal part, and
$\mathbf{\zeta}_t(\mathbf{x},t)$ the transverse part,
\begin{equation}\label{rotul}
\nabla\times \mathbf{\zeta}_l=0~,\hspace{5mm}\nabla \cdot
\mathbf{\zeta}_t=0~.
\end{equation}
One can then obtain the wave equations with dissipation terms
\cite{Royer,Stokes,Buckingham},
\begin{equation}\label{basicut}
\ddot{\mathbf{\zeta}}_t-c_t^2 \triangle
\mathbf{\zeta}_t-\frac{1}{\rho}\eta_1 \triangle \dot{\mathbf{\zeta}}_t=0~,
\end{equation}
\begin{equation}\label{basicul}
\ddot{\mathbf{\zeta}}_l-c_l^2 \triangle
\mathbf{\zeta}_l-\frac{1}{\rho}\left(2\eta_1+\eta_2\right)\triangle
\dot{\mathbf{\zeta}}_l = 0~.
\end{equation}
In these expressions $c_{l,t}$ are the velocities of
longitudinal/transverse sound waves, $c_t = \sqrt{\lambda_1/\rho}$, $c_l = \sqrt{(2\lambda_1+\lambda_2)/\rho}$. We switch from time-- to frequency domain by means of
\begin{equation}\label{u}
\mathbf{\zeta} = \mathbf{u}e^{i \tilde{\omega} t}~,~\epsilon = \tilde{\epsilon}~e^{i\tilde{\omega}t}~,
\end{equation}
where we always take the real part. The frequency $\tilde{\omega}$ for non--vanishing viscosity takes complex values with positive imaginary part. All real valued frequencies in this paper we denote generally without a tilde. 
The vector $\mathbf{u}$ has the components $\{u_r,u_\theta,u_\phi\}$,
\begin{equation}\label{udef}
\mathbf{u} = u_r \mathbf{e}_r + u_\theta\mathbf{e}_\theta + u_\phi \mathbf{e}_\phi~, 
\end{equation}
and we obtain the Helmholtz equations with complex wave numbers $k_{t}$, $k_l$,
\begin{equation}\label{utomega}
\triangle
\mathbf{u}_t + k_t^2\mathbf{u}_t=0~,\hspace{5mm}k_t = \frac{\tilde{\omega}}{\sqrt{c_t^2+i\frac{\tilde{\omega} \eta_1}{\rho }}}~,
\end{equation}
\begin{equation}\label{ulomega}
\triangle
\mathbf{u}_l + k_l^2\mathbf{u}_l = 0~,\hspace{5mm}k_l = \frac{\tilde{\omega}}{\sqrt{c_l^2+i\frac{\tilde{\omega}}{\rho}\left(2\eta_1+\eta_2\right)}}~.
\end{equation}
We furthermore define the complex viscoelastic moduli
\begin{equation}\label{complexlambda}
~\overline{\lambda}_1: = \lambda_1 + i \tilde{\omega} \eta_1~,~\overline{\lambda}_2: = \lambda_2 + i \tilde{\omega} \eta_2~,
\end{equation}
and the dimensionless wave numbers
\begin{equation}
R_t:=R k_t~,~ R_l:=R k_l~.
\end{equation}
Similar to Eq.~\eqref{u}, we define 
\begin{equation}\label{tildeT1}
\delta T(\mathbf{x},t) = \delta \tilde{T}(\mathbf{x})e^{i\tilde{\omega} t} ~,
\end{equation}
allowing for treatment of the heat equation.
\subsection{Frequency shifts}
To calculate the frequency shifts caused by viscosity, we can use only the approach of the zeros in the complex plane. The approach Eq.~\eqref{gdissint} does not allow a prediction of frequency shifts. In order to gain more insights, the two limiting cases  of longitudinal/transverse modes are of particular interest. 
Using Eq.~\eqref{ulomega} one obtains for the one dimensional, longitudinal case, for the real part $\omega$ of the complex frequency,
\begin{equation}\label{shiftlong}
\omega=\omega_0\sqrt{1-\Big(\frac{2\eta_1+\eta_2}{2\rho c_l}\Big)^2\frac{\omega_0^2}{c_l^2}}<\omega_0~,
\end{equation}   
where $\omega_0$ corresponds to the frequency without damping and carrying the information of the boundary conditions. Similarly, for the one dimensional, transverse case, Eq.~\eqref{utomega}
\begin{equation}\label{shifttrans}
\omega=\omega_0\sqrt{1-\Big(\frac{\eta_1}{2\rho c_t}\Big)^2\frac{\omega_0^2}{c_t^2}}<\omega_0~.
\end{equation}   
Both expressions show a decrease of the resonance frequencies when viscosity is present, i.e. $\omega<\omega_0$, as one would expect. If we consider our problem, we have to work again with the functions $f(\omega+i\gamma;\eta_1,\eta_2)$. 
We use a Taylor expansion, now at the point in the complex plane $\omega_0+i\gamma_{\mathrm{diss,visc}}$, which yields
\begin{equation}\label{shifttaylor}
\omega = \omega_0 - \mathrm{Re}\Bigg[\frac{f(Z)}{\Big(\frac{\partial f}{\partial\omega}\Big)_Z}\Bigg]~,
\end{equation}
with the argument triple $Z:=(\omega_0+i\gamma_{\mathrm{diss,visc}},\eta_1,\eta_2)$. Here, however, the frequency shift $\frac{\omega_0-\omega}{\omega_0}$ we obtain are not necessarily positive. The frequency shift caused by thermal conduction is not considered in this work. A detailed description can be develop on the basis of the derived temperature fields \cite{PhysRevB.61.5600}.
\section{Solutions for the different vibration modes}\label{sect2}
The solution steps for the elastic problem are well known and reported in the literature \cite{Eringen}. We re--derive the expressions, including the viscosity tensor. Because we are also interested in the interplay between displacement and temperature, we use a dimensionless real parameter $\mathcal{A}$ to increase/decrease the amplitude of the displacement. This also affects the temperature with the same factor of $\mathcal{A}$. 
\subsection{Breathing modes}
We begin with the breathing modes (spherically symmetric pulsations), which are a special case of the more general spheroidal modes with $n=0$ and $m=0$. The condition for the eigenfrequencies reads 
\begin{equation}\label{Lambq}
f_0(\tilde{\omega};\eta_1,\eta_2) = 4\overline{\lambda}_1 R_l\cos R_l + \Big(2\overline{\lambda}_1(R_l^2-2)+\overline{\lambda}_2 R_l^2\Big)\sin R_l \overset{!}{=}  0~,
\end{equation}
and for the displacement we formulate the expression
\begin{equation}\label{ubreath}
\mathbf{u} = -\mathcal{A}R^2k_l\{j_1(k_l r),0,0\}~,
\end{equation}
where $j_l(x)$ is the spherical Bessel function of order $l$ \cite{Abramowitz}. This agrees with  \cite{Lamb} for vanishing dissipation, i.e. for $\overline{\lambda}_1 = \lambda_1$ and $\overline{\lambda}_2 = \lambda_2$. In this (dissipation less) case Eq.~(\ref{Lambq}) asymptotically gives zeros for $\omega_k = \frac{c_l}{R}k~ \pi~,~k\gg 1$. Next we evaluate the integrals Eq.~\eqref{gV} and Eq.~\eqref{gT}. Let $\omega_k$, $k=1,2,3...$, be the roots of Eq.~\eqref{Lambq} without damping ($\eta_1=\eta_2=0$) and $S_k := \omega_k R/c_l$. For the damping of the $k$-th mode we obtain
\begin{eqnarray}\label{gVBM}
\gamma_{\mathrm{diss,visc}}=
\frac{\eta_1\Big(8+8 S_k^2-4S_k^4+8(S_k^2-1)\cos(2 S_k) + 2 S_k(S_k^2-8)\sin(2 S_k)\Big) - \eta_2\Big(2S_k^4-S_k^3\sin(2S_k)\Big)}{2R^2\rho\Big(2-2S_k^2-2\cos(2S_k) -S_k\sin( 2S_k)\Big)}~,
\end{eqnarray}
and
\begin{equation}\label{gTbrm}
\gamma_{\mathrm{diss,T}} = \frac{\mathrm{k} T_v^2\omega_k^2}{2T_0\rho c_l^4}~.
\end{equation}
Eq.~\eqref{gVBM} shows the asymptotic behavior (keeping $\omega_k$ constant; thermodynamic limit)
\begin{equation}\label{limitgbreath}
\lim_{R\rightarrow\infty}\gamma_{\mathrm{diss,visc}}=\frac{(2\eta_1+\eta_2)\omega_k^2}{2\rho c_l^2}~,
\end{equation}
which, up to a factor $1/c_l$, agrees with the corresponding expression for the attenuation of sound, caused by viscosity, derived for plane waves in an infinite space in \cite{Landau7}. The same holds true for $\gamma_{\mathrm{diss,T}}$. For completeness we also provide the result for $\gamma_{\mathrm{diss,visc}}$ according to Eq.~\eqref{gValternativ},
\begin{eqnarray}\label{gVBMalt}
\gamma_{\mathrm{diss,visc}}= \Bigg\{\eta _1 \Big(4  \left(2 \rho  c_l^2+\lambda _2 S_k^2 -\rho 
   c_l^2 S_k^2 \right)\sin \left(S_k\right) + 2\rho  c_l^2   \left(S_k^3- 4 S_k\right) \cos \left(S_k\right)  \Big) +\nonumber{}\\
 \eta _2 S_k^2 \Big(2\left(\lambda _2-\rho  c_l^2\right)\sin \left(S_k\right)+\rho  c_l^2 S_k \cos \left(S_k\right)\Big)\Bigg\}\Bigg/\nonumber{}\\
\Bigg\{ 2 \rho  R^2 \Big(\rho  c_l^2 S_k \cos
   \left(S_k\right)+2 \lambda _2 \sin \left(S_k\right)\Big)\Bigg\}~.
\end{eqnarray}
This expression possesses the same limit as Eq.~\eqref{limitgbreath}.
For the temperature distribution we obtain by direct integration of the heat equation Eq.~\eqref{basicT} without the angular part in the Laplacian and using $\omega_k$ on the r.h.s. for the displacement field, 
\begin{equation}\label{tildeTbr}
\delta \tilde{T}(r) = i \mathcal{A}R^2 k_l T_v\frac{\omega_k}{\alpha}\frac{\Big(S_k\cos(S_k) - \sin(S_k)\Big)\sin(\nu_k r) + \Big(\sin(\nu_k R)-R\nu_k \cos(\nu_k R)\Big)\sin(\omega_k r/c_l)}{r\Big(\frac{\omega_k^2}{c_l^2}-\nu_k^2\Big)\Big(\nu_k R \cos(\nu_k R) - \sin(\nu_k R)\Big)}~.
\end{equation}
Here we have defined $\nu_k := \sqrt{-i \omega_k/\alpha}$. Note that $\delta \tilde{T}(r)$ is characterized by the single number $k$, the $k$-th zero of Eq.~\eqref{Lambq}. This expression is valid also for the complex frequency $\tilde{\omega}_k$.
\subsection{Spheroidal modes}
The corresponding factor of the general determinant after simplification reads (see our accompanying Mathematica notebook \textit{DampingViscoelasticSphere.nb} for further details),
\begin{eqnarray}\label{omegasph}
f_n(\tilde{\omega};\eta_1,\eta_2) = 4 R_l J_{n+\frac{3}{2}}(R_l)\Bigg\{(n^3+2n^2-n-2 - R_t^2)J_{n+\frac{1}{2}}(R_t) - (n^2+n-2)R_t J_{n+\frac{3}{2}}(R_t)\Bigg\} + \nonumber{}\\
J_{n+\frac{1}{2}}(R_l)\Bigg\{\Big[2n(1-n)R_t^2 - R_l^2(2n^2-2-R_t^2)\frac{2\overline{\lambda}_1+\overline{\lambda}_2}{\overline{\lambda}_1}\Big]J_{n+\frac{1}{2}}(R_t)+\nonumber{}\\
2R_t\Big[2n^3+2n^2-4n-\frac{2\overline{\lambda}_1+\overline{\lambda}_2}{\overline{\lambda}_1} R_l^2\Big]J_{n+\frac{3}{2}}(R_t)
\Bigg\}\overset{!}{=}  0~.
\end{eqnarray}
Here $J_{n+\frac{3}{2}}(x)$ is a Bessel function of order $n+3/2$. To derive the displacement field from the general expressions (see App.~\ref{appa}), we have to calculate $A_l$, $B_t$ and $C_t$. We can put $A_l=1$, and obtain $B_t=0$. With $n=1,2,3,\hdots,\ -n\leq m\leq n$, we find

\begin{eqnarray}\label{urspher}
u_r = \mathcal{A}\frac{R^2 e^{i m \phi }}{r}  P_n^m(\cos \theta)\Big[n j_n(k_l r)+n(1+n)C_t j_n(k_t r) - k_l r j_{n+1}(k_l r)\Big]~,
\end{eqnarray}

\begin{eqnarray}\label{uthspher}
u_\theta = -\mathcal{A}\frac{R^2e^{i m \phi }}{r} \csc \theta  \Big[(n+1) \cos \theta P_n^m(\cos \theta)+(m-n-1) P_{n+1}^m(\cos \theta
  )\Big]  \times\nonumber{}\\
\Big[C_t (n+1) j_n(k_t r)-C_t k_t r j_{n+1}(k_t r)+j_n(k_l r)\Big]~,
\end{eqnarray}
\begin{equation}\label{uphspher}
u_\phi=i\mathcal{A}\frac{ m R^2 e^{i m \phi }}{r} \csc \theta P_n^m(\cos\theta) \Big[C_t (n+1) j_n(k_t r)-C_t
   k_t r j_{n+1}(k_t r)+j_n(k_l r)\Big]~,
\end{equation}
with $P^m_n(x)$ being the associated Legendre polynomials. The constant $C_t$ is given by
\begin{eqnarray}\label{Ct}
C_t = \frac{(1-n)j_n(R_l) + R_l j_{n+1}(R_l)}{(n^2-1-\frac{1}{2}R_t^2)j_n(R_t)+R_t j_{n+1}(R_t)}~.
\end{eqnarray}
The integrals Eq.~\eqref{gV} and Eq.~\eqref{gT} in this case are not analytically available in a general form, and so we can use numerical quadrature techniques instead. The evaluation of Eq.~\eqref{gValternativ} for Eq.~\eqref{omegasph} on the other hand is still possible, yielding an extensive expression. In fact, this is the most interesting application of Eq.~\eqref{gValternativ} in this work. To make the result more clear, we write it in the following form (here $\omega_k$ denotes a zero of Eq.~\eqref{omegasph} with zero damping),
\begin{equation}
\gamma_{\mathrm{diss,visc}}=\frac{\omega_k^2}{2\rho c_l^2}\Big[(2\eta_1+\eta_2)\psi_L(n,\omega_k)+\eta_1\psi_T(n,\omega_k)\Big]~.
\end{equation}
The two auxiliary functions $\psi_L(n,\omega)$ and $\psi_T(n,\omega)$ are defined in Appendix \ref{appb}. Asymptotically, $f_n(\tilde{\omega};\eta_1,\eta_2)$, $R\rightarrow\infty$, $k\rightarrow\infty$, $\eta_1=0$, $\eta_2=0$ (dissipation less case), has two kinds of zeros. We follow the concept discussed in \cite{PhysRevB.72.205433} and call these mathematically not exact, distinct modes `\textit{primarily longitudinal}' and `\textit{primarily transverse}' modes.
The two series of zeros in Eq.~\eqref{omegasph} have consequently two limiting cases for the damping constant. Let us define two series of frequencies, $\Omega_{l,k} :=\frac{c_l\pi}{R}\left(\frac{n}{2} +k\right)$, and $\Omega_{t,k} :=\frac{c_t\pi}{R}\left(\frac{n}{2} +k\right)$, $k=1,2,3,\hdots$. For primarily longitudinal modes we then have 
\begin{equation}\label{omegalong}
\omega_k \rightarrow \Omega_{l,k}~,~\psi_L(n,\omega_k)\rightarrow 1~,~\psi_T(n,\omega_k)\rightarrow 0~,\gamma_{\mathrm{diss,visc}}\rightarrow\frac{(2\eta_1+\eta_2)\omega_k^2}{2\rho c_l^2}~,
\end{equation} 
and for primarily transverse modes
\begin{equation}\label{omegatransv}
\omega_k \rightarrow \Omega_{t,k}~,~\psi_L(n,\omega_k)\rightarrow 0~,~\psi_T(n,\omega_k)\rightarrow \frac{c_l^2}{c_t^2}~,\gamma_{\mathrm{diss,visc}}\rightarrow\frac{\eta_1\omega_k^2}{2\rho c_t^2}~.
\end{equation} 
The damping $\gamma_{\mathrm{diss,T}}$ is computable analytically using Eq.~\eqref{gT}, if $n$ and $m$ take concrete numbers. An example expression is given in our Mathematica notebook. To investigate the two limiting cases it is useful to consider the constant $C_t$ through Eq.~\eqref{Ct} first, which approaches the values $0$ or $\infty$. Conducting this calculation for the first few values, we obtain for primarily longitudinal modes in accordance with Eq.~\eqref{gTbrm},
\begin{equation}\label{omegalongT}
\omega_k \rightarrow \Omega_{l,k}~,~C_t\rightarrow 0~,~\gamma_{\mathrm{diss,T}}\rightarrow\frac{\mathrm{k} T_v^2\omega_k^2}{2T_0\rho c_l^4}~,
\end{equation} 
and for primarily transverse modes (corresponding to Eq.~\eqref{gTtor}) we find
\begin{equation}\label{omegatransvT}
\omega_k \rightarrow \Omega_{t,k}~,~C_t\rightarrow \infty~,~\gamma_{\mathrm{diss,T}}\rightarrow 0~.
\end{equation} 
Even if this jumping behavior (see left panel of Fig.~\ref{fig3}) seems reasonable for all $n$ and $m$ values, we have no exact proof of this behavior (for the primarily transverse modes the vanishing divergence ensures the limit $\gamma_{\mathrm{diss,T}}\rightarrow 0$). We now derive the temperature field. First we denote the divergence of the displacement field,
\begin{equation}\label{div}
\nabla\cdot \mathbf{u} = -\mathcal{A} R^2 k_l^2 j_n(k_l r)P^m_n(\cos\theta)e^{i m\phi}~.
\end{equation}
This expression suggests the following ansatz,
\begin{equation}\label{tildeT1cyl}
\delta \tilde{T}(r,\theta,\phi) = -i\mathcal{A} R^2\omega_k T_v P^m_{n}(\cos\theta)e^{i m \phi}\sum_{\nu=1}^{\infty}\tilde{T}_\nu j_n(\hat{k}_{n,\nu}r)~,
\end{equation}
with a priori unknown series coefficients $\tilde{T}_{\nu}$. Because we assume an adiabatic boundary condition at the surface, the $\hat{k}_{n,\nu}$ are given by
\begin{equation}
\Big(\frac{d j_n(x)}{d x}\Big)_{x=\hat{k}_{n,\nu}R}=0~,
\end{equation}
where $\nu$ enumerates the $\nu$-th zero. To obtain the coefficients $\tilde{T}_\nu$ we start with the eigenvalue equation $\triangle j_n(\hat{k}_{n,\nu}r) P^m_n(\cos\theta)e^{i m \phi}=-\hat{k}_{n,\nu}^2 j_n(\hat{k}_{n,\nu}r) P^m_n(\cos\theta)e^{i m \phi}$. We use orthogonality, valid for arbitrary $n$ \cite{Hilbert},
\begin{equation}\label{norm}
\int_0^{R} r^2 j_n(\hat{k}_{n,\nu}r)j_n(\hat{k}_{n,\nu'}r) dr = \delta_{\nu\nu'}N_{n,\nu}~,
\end{equation}
where $\delta_{nn'}$ is the Kronecker symbol, and with the normalization
\begin{equation}\label{norm1}
N_{n,\nu} = \int_0^{R} r^2 j_n(\hat{k}_{n,\nu}r)^2 dr = \frac{R^3}{2}\Big(j_n(\hat{k}_{n,\nu}R)^2-j_{n-1}(\hat{k}_{n,\nu}R)j_{n+1}(\hat{k}_{n,\nu}R)\Big)~.
\end{equation}
Furthermore, we have to calculate the overlap integral with $j_n(k_l r)$ from Eq.~\eqref{div},
\begin{equation}\label{eqM}
M_{n,\nu} = \int_0^{R} r^2 j_n(\hat{k}_{n,\nu}r)j_n(k_l r) dr = \frac{R^2}{k_l^2-\hat{k}_{n,\nu}^2}\Big(\hat{k}_{n,\nu}j_{n-1}(\hat{k}_{n,\nu}R)j_{n}(k_{l}R)-k_l j_{n-1}(k_{l}R)j_{n}(\hat{k}_{n,\nu}R)\Big)~.
\end{equation}
We arrive at a fast converging series for the temperature, 
\begin{equation}\label{tildeT2cyl}
\delta \tilde{T}(r,\theta,\phi) = -i\mathcal{A}R^2 k_l^2\omega_k T_v P^m_{n}(\cos\theta)e^{i m \phi}\sum_{\nu=1}^{\infty}\frac{M_{n,\nu}j_n(\hat{k}_{n,\nu}r)}{(\alpha\hat{k}_{n,\nu}^2+i\omega_k)N_{n,\nu}}~.
\end{equation}
Note that in this case $\delta \tilde{T}(r,\theta,\phi)$ is characterized by three numbers, the mode numbers $m,n$ and $k$, tied to the enumeration of zeros of Eq.~\eqref{omegasph}. As denoted for the solution Eq.~\eqref{tildeTbr}, it is possible also here to use complex frequencies for the r.h.s. in Eq.~\eqref{basicT}.
\subsection{Torsional modes}\label{sect3}
In the case of torsional modes the function $f_n(\tilde{\omega};\eta_1)$ is given by ($n=1,2,3,\hdots$, $-n\leq m\leq n$),
\begin{eqnarray}\label{omegators}
f_n(\tilde{\omega};\eta_1) = (n-1)j_{n}(R_t) - R_t j_{n+1}(R_t)\overset{!}{=}  0~.
\end{eqnarray}
With $A_l=0$, $C_t=0$ and $B_t=1$, we find a displacement of
\begin{eqnarray}\label{urtors}
u_r = 0 ~,
\end{eqnarray}

\begin{eqnarray}\label{uthtors}
u_\theta = i\mathcal{A}R \frac{m e^{i m\phi}}{\sin\theta}P^m_n(\cos\theta)j_n(k_t r)~,
\end{eqnarray}

\begin{eqnarray}\label{uphtors}
u_\phi=\mathcal{A}R\frac{e^{i m\phi}}{\sin\theta} 
\Big[(1+n)\cos\theta P^m_n(\cos\theta) + (m-n-1)P^m_{n+1}(\cos\theta)\Big]j_n(k_t r)
~.
\end{eqnarray}
These expressions agree with \cite{HOSSEINIHASHEMI1988511} in the absence of dissipation, where one can find the normalization and orthogonality properties. 
Since $\nabla\cdot \mathbf{u}=0$ for this displacement field, we conclude from Eq.~\eqref{gT}
\begin{equation}\label{gTtor}
\gamma_{\mathrm{diss,T}}=0~,
\end{equation}
as well as 
\begin{equation}
\delta \tilde{T}(r,\theta,\phi)=0~.
\end{equation}
Any small temperature change for torsional modes is therefore caused by viscosity only.
With the help of Mathematica we obtain analytical expressions for $\gamma_{\mathrm{diss,visc}}$, when $n$ is given. As an example we give here the results for the first two modes, where we define $P_k:=\omega_k R/c_t$.
For $n=1, m = -1,0,1$, we have 
\begin{eqnarray}\label{gVTM1}
\gamma_{\mathrm{diss,visc}}=
-\frac{\eta_1 \Big(-2 P_k^4+6 P_k^2+\left(P_k^2-12\right) P_k \sin (2
   P_k)+6 \left(P_k^2-1\right) \cos (2 P_k)+6\Big)}{2 \rho R^2 \Big(2
   P_k^2+P_k \sin (2 P_k)+2 \cos (2 P_k)-2\Big)}~.
\end{eqnarray}
For $n=2, m = -2,-1,0,1,2$, we find 
\begin{eqnarray}\label{gVTM2}
\gamma_{\mathrm{diss,visc}}=
\frac{\eta_1  \Big(P_k \left(P_k^4-48 P_k^2+144\right) \sin (2 P_k)+2 \left(5
   P_k^4-60 P_k^2+36\right) \cos (2 P_k)+2 \left(P_k^6-5 P_k^4-12
   P_k^2-36\right)\Big)}{2 \rho R^2 \Big(2 P_k^4-6
   P_k^2-\left(P_k^2-12\right) P_k \sin (2 P_k)+\left(6-6 P_k^2\right) \cos (2
   P_k)-6\Big)}.
\end{eqnarray}
The two examples (and also some higher mode numbers $n$ investigated) show no dependence on $m$, as one would expect from Eq.~\eqref{omegators}, which also does not depend on $m$, and for which the imaginary part of the zeros is related to $\gamma_{\mathrm{diss,visc}}$. Let us consider the limits for Eqs.~(\ref{gVTM1},\ref{gVTM2}), 
\begin{equation}\label{gtors}
\lim_{R\rightarrow\infty}\gamma_{\mathrm{diss,visc}}=\frac{\eta_1\omega_k^2}{2\rho c_t^2}~.
\end{equation}
Interestingly, Eq.~\eqref{gValternativ} immediately gives this result.
\section{Numerical example and discussion}\label{sect4}
\begin{figure}
\begin{center}
	\includegraphics[height=55 mm, angle=0]{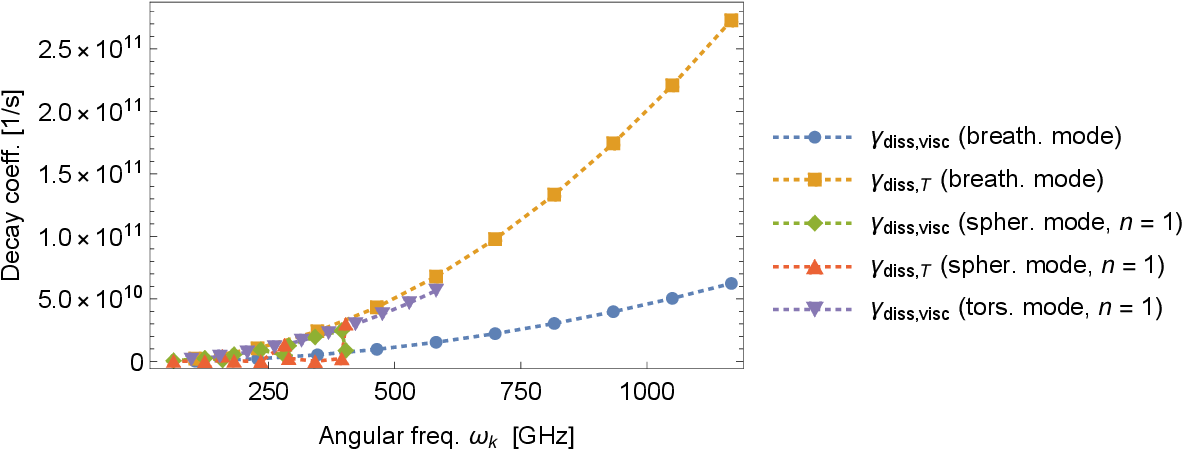}
\caption{\label{fig1}(Color online) Decay coefficients for the first ten frequencies for the fundamental eigenmodes.}
\end{center}
\end{figure}

\begin{figure}
\begin{center}
\includegraphics[height=55 mm, angle=0]{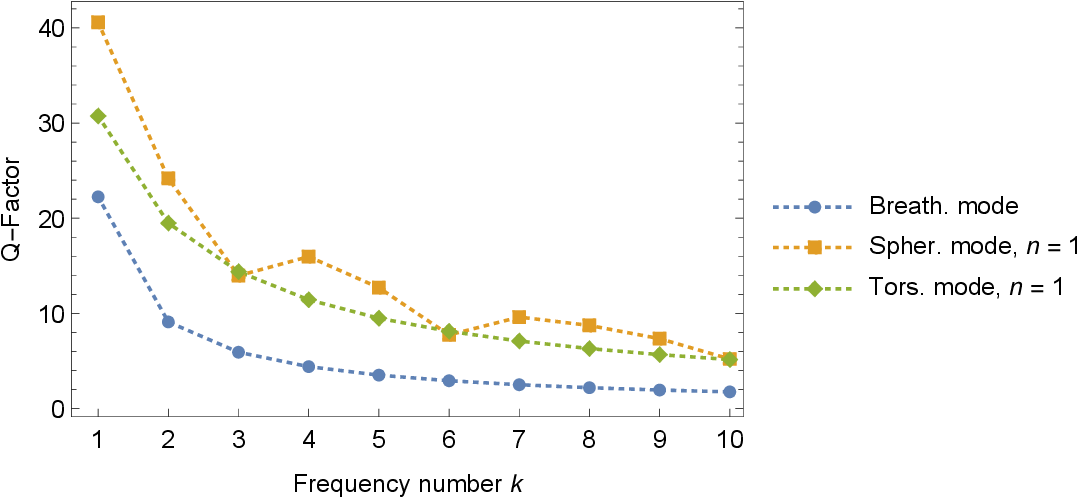}
\caption{\label{fig2}(Color online) Quality factor $Q :=\omega_k/2\gamma_{\mathrm{diss,tot}}$ for the first ten eigenfrequencies.}
\end{center}
\end{figure}

\begin{figure}
\begin{center}
\includegraphics[height=100 mm, angle=0]{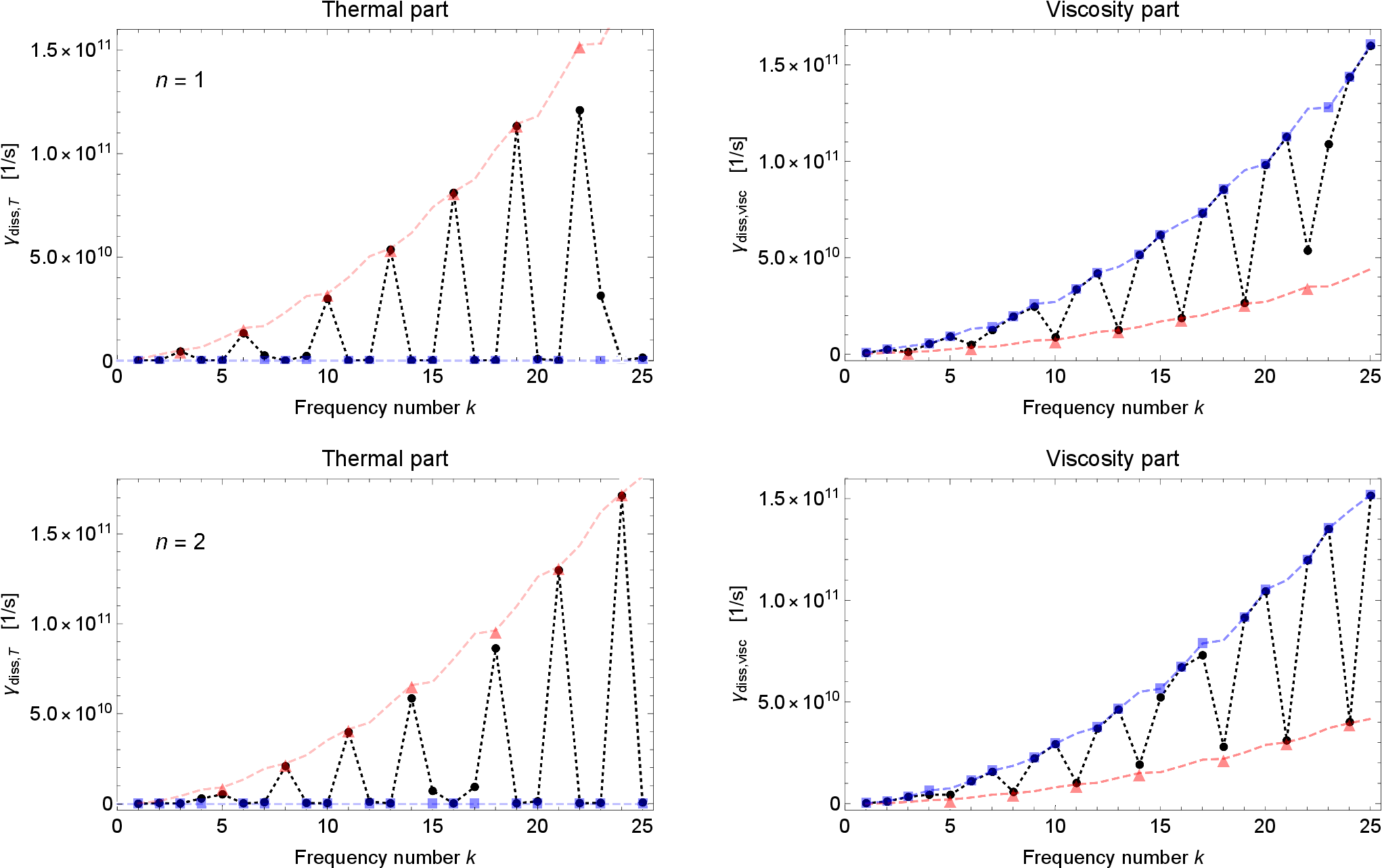}
	\caption{\label{fig3}(Color online) Damping constants $\gamma_{\mathrm{diss,T}}$, $\gamma_{\mathrm{diss,visc}}$ for spheroidal modes with $n=1$ (upper row)  and $n=2$ (lower row). The black dots are the results from exact calculations. The red triangles show the limiting values for primarily longitudinal vibrations, and the blue squares for primarily transverse vibrations. For the first modes the distinction is not clear and somewhat arbitrary, for higher frequencies, however, the difference becomes fully manifest.}
\end{center}
\end{figure}

In this section we give an impression of the results obtained in the previous sections. We consider a viscoelastic Ag sphere with  a radius of $R=100$~nm and use the material parameters listed in Tab.~I. It is known that the elastic properties show a significant spread and vary across the literature \cite{Kohlrausch}. Note furthermore, that we use the elastic values for bulk materials and static stresses. For the viscosity coefficients, the situation is even more difficult (see discussion in \cite{PhysRevLett.82.1478}). We base our investigation on Stokes hypothesis $\zeta_V=0$, or $\eta_2 = -2\eta_1/3$ for our simulations,  \cite{Bosworth,PhysRev.173.856}. To approximate a realistic value, compare to the experimental values (quality factor) for an aluminium sphere with radius 153 mm \cite{COCCIA1996263}.
Fig.~\ref{fig1} shows the damping for the various modes for the first ten eigenfrequencies. The different values for $\gamma_{\mathrm{diss,visc}}$, $ \gamma_{\mathrm{diss,T}}$ were obtained by the corresponding analytical expressions or the numerical evaluation of Eq.~\eqref{gV} and Eq.~\eqref{gT}. We also solved the eigenfrequency equations for the complex eigenfrequencies $\tilde{\omega}_k = \mathrm{Re}(\tilde{\omega}_k) + i~\mathrm{Im}(\tilde{\omega}_k)$. For the chosen $\eta_{1,2}$ the values are generally in good agreement with one another, $\gamma_{\mathrm{diss,visc}} = \mathrm{Im}(\tilde{\omega}_k)$ and also with the ones from Eq.~\eqref{gValternativ}. For the spheroidal modes, when evaluating the integral Eq.~\eqref{gV} with the displacement field  Eqs.~(\ref{urspher}--\ref{uphspher}), we find no dependence on the magnetic quantum number $m$.\\
In Fig.~\ref{fig2} we plot the quality factor $Q$ for the various modes. While the breathing and torsional modes show a monotonically decreasing behavior, the spheroidal modes are non--monotonic, indicating the `primarily longitudinal' and `primarily transverse' character of the vibrations. $Q$ depends significantly on the sphere's radius: an increase of the sphere's radius by a factor $g$ decreases the frequencies by a factor of $1/g$, and increases the quality factor by $g$ (nearly exact, because for the limiting cases $Q\propto R$).\\
Fig.~\ref{fig3} shows the damping constants for the spheroidal modes ($n=2$), where the thermal-- and viscosity parts are plotted separately for the first 25 eigenfrequencies. As before, we observe the non--monotonic behavior in both $\gamma_{\mathrm{diss,T}}$, $\gamma_{\mathrm{diss,visc}}$, as discussed in the previous section.
We have plotted also the limiting cases Eq.~\eqref{gTbrm}, Eq.~\eqref{gTtor}, Eq.~\eqref{limitgbreath} and Eq.~\eqref{gtors}, which envelope the actual values. 
The plot shows two features: (i) the existence of finite size effects, and (ii) the concept of `primarily longitudinal' and `primarily transverse' modes is valuable/applicable also for damping constants, when higher frequencies are considered. We emphasize that even in this case the three approaches of the calculation of complex roots in Eq.~\eqref{omegasph}, the evaluation of the integrals Eq.~\eqref{gV}, and the evaluation of the Taylor formula, Eq.~\eqref{gValternativ}, for the chosen parameters generally lead to similar numerical values.\\

\begin{figure}
\begin{center}
\includegraphics[height=50 mm, angle=0]{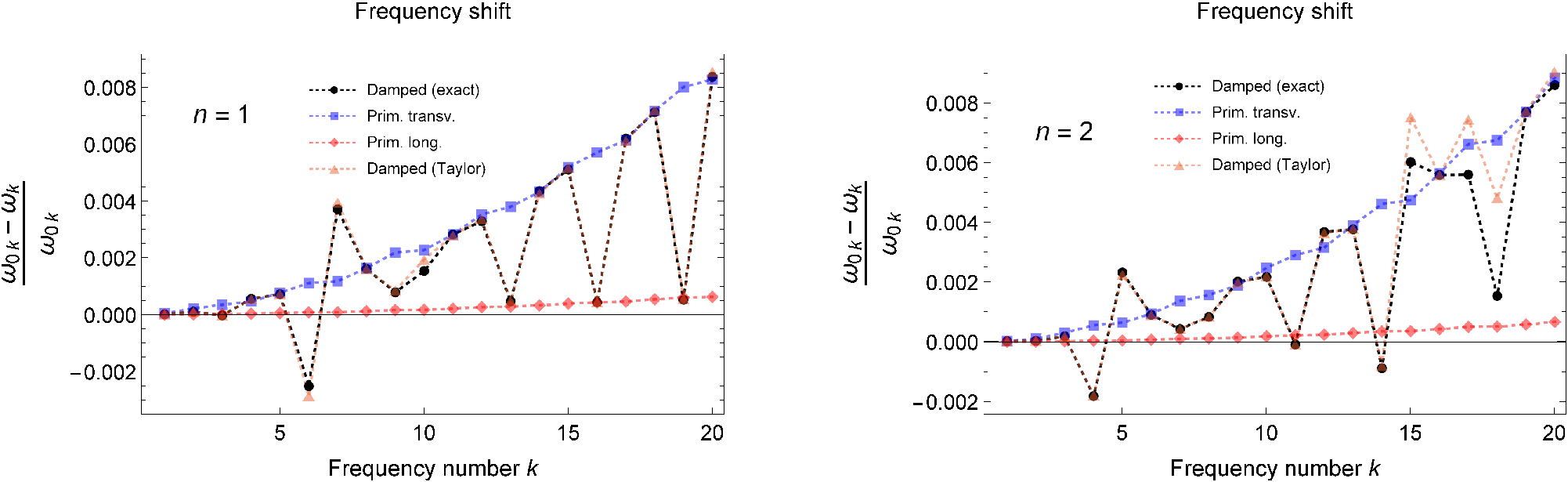}
	\caption{\label{fig4}(Color online) Frequency shift of the first 20 frequencies for spheroidal modes with $n=1$ (left) and $n=2$ (right). Additional to the exact values (black dots) we have plotted also the analytical values according to Eq.~\eqref{shiftlong}, Eq.~\eqref{shifttrans} and Eq.~\eqref{shifttaylor}.}
\end{center}
\end{figure}

\begin{figure}
\begin{center}
\includegraphics[height=50 mm, angle=0]{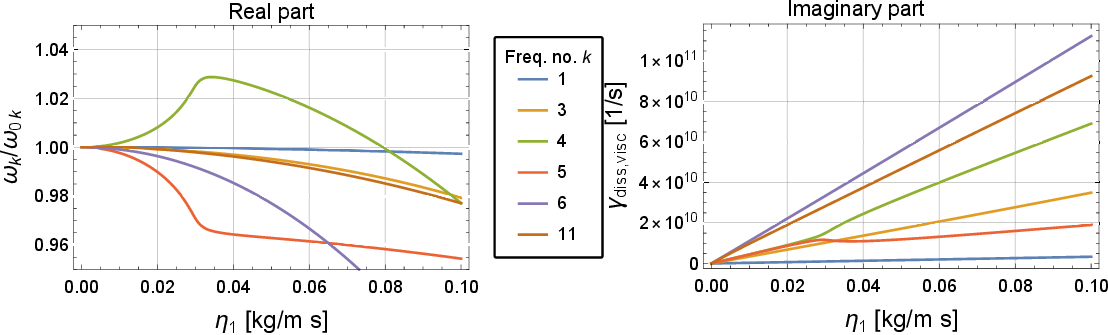}
\caption{\label{fig5}(Color online) The left part of the figure shows the frequency ratio damped/undamped, $\omega_k/\omega_{0k}$ for selected eigenfrequencies as a function of viscosity $\eta_1$ (with $\eta_2=-2\eta_1/3$) for spheroidal modes with $n=2$. One can see that the fourth frequency shows a ratio $\omega_k/\omega_{0k}>1$ (the maximum is $1.0287$). For the 11. frequency this behavior is restricted to a small, on this scale not visible $\eta_1$ region. The right part of the figure shows the damping, calculated from the imaginary part of the complex zeros. The fourth and fifth frequencies show some anomaly compared with the others. A similar behavior we have found in the case $n=5$ for the fifth/sixth frequency pair.}
\end{center}
\end{figure}

Let us now discuss the frequency shifts. For the breathing/torsional modes the theory gives results as expected from Eq.~\eqref{shiftlong}, Eq.~\eqref{shifttrans}. A table with numerical results is given in our Notebook. 
More interesting are the spheroidal modes, given by the zeros in the complex plane of Eq.~\eqref{omegasph}.  
Fig.~\ref{fig4} shows the shifts for $n=2$ and the first 25 eigenfrequencies. The lower (4th and 5th) frequencies are hard to understand in its physical meaning. The higher frequencies behave in accordance with the imaginary part $\gamma_{\mathrm{diss,visc}}$ and the concept of `primarily longitudinal' and `primarily transverse' modes, now given through Eq.~\eqref{shiftlong}, Eq.~\eqref{shifttrans}. The 4th mode however is of particular interest and needs a more detailed discussion. To illuminate the unexpected shift of the fourth frequency, we have changed the viscosity, whereas all other parameters remained constant. In Fig.~\ref{fig5} we have plotted some selected frequency ratios damped/undamped versus $\eta_1$, and we see that for the 4th frequency  $\omega_k/\omega_{0k}>1$ in the range $0\leq \eta_1 <\approx 0.08$, meaning the counter-intuitive behavior that an increase of viscosity leads to an increase of the resonance frequency, \footnote{To be safe that this result is correct, we checked numerically the eigenvalue equation $\bar{\lambda}_1\triangle \bf{u}+(\bar{\lambda}_1+\bar{\lambda}_{2})\nabla(\nabla\cdot \bf{u})+\rho\tilde{\omega}^{2}\bf{u}=0$ on a grid inside the sphere and the boundary condition with high accuracy. The complex frequencies itself can be find convenient, starting at the real axis and increasing step-wise, with small increment, the parameters $\eta_{1,2}$ such that the last result can serve as initial guess for the solver for the next root.}. The introduction of a non--vanishing volume viscosity $\eta_2+2\eta_1/3$ does not cure this problem. Also the frequencies $\omega_{11}$ and $\omega_{14}$ show this anomaly to a weaker extent. However, when increasing the radius of the sphere by a factor of $g$, the maximum of the ratio of $\omega_4/\omega_{04}$ remains constant and equal to $\omega_4/\omega_{04}=1.0287$, whereas the position of this maximum (viscosity) increases by the same factor of $g$.
If these anomalies have physical relevance, we could explain it by the fact that the spheroidal modes are not purely longitudinal or transverse vibrations. But these results are also of mathematical interest, because we expected that the viscosity tensor Eq.~\eqref{sigmas} ensures not only positive imaginary parts in the frequencies (which indeed is the case), but also frequency ratios smaller than one, which seems to not be the case here. It is interesting that only very specific modes show this particular behavior, and further investigations will be necessary to gain deeper insights into the mechanisms behind this phenomenon. 

\begin{table}
\begin{center}
\begin{tabular}{|c|c|c|c|c|c|}
  \hline
  $\rho$~[kg/m$^3$] & $\lambda_1$~[GPa] & $\lambda_2$~[GPa] & $K$~[GPa] & $\alpha_T$~[1/K] & k~[W/mK]\\
  \hline
  10490 & 29.9 & 85.1 & 105 & $5.85\times 10^{-5}$ & 429 \\
\hline \hline $c_l$~[m/s] & $c_t$~[m/s] & $\eta_1$~[kg/m~s]&
$\eta_2$~[kg/m~s] & $\alpha$ [m$^2$/s] & $c_p$~[J/kgK] \\ \hline
 3718 & 1689 & 0.01  &-0.00667 & $1.76 \times 10^{-4}$& 232\\   \hline
\end{tabular}
\caption{Parameters for Ag, used in the numerical simulations. The
values for $\eta_1$ and $\eta_2$ are estimates. Circular frequencies for spheres with radii of about 100 nm are in the range of around $>$100 GHz.}\label{tab1}
\end{center}
\end{table}

\section{Summary and Outlook}
In this paper we have investigated the damping of a homogeneous (nano)sphere using analytical methods for solving the visco--elastic problem and the heat equation. We present a collection of analytical formulas also suitable for numerical computations. General formulas, such as Eq.~\eqref{gV} and Eq.~\eqref{gT}, are derived, and exact results are given for breathing modes and torsional modes. For spheroidal modes we use an alternative approach (Eq.~\eqref{gValternativ}), and derive an expression for the damping constant. We find non--monotonic behavior for these modes, corresponding to the presence of two kinds of vibration modes, characterized by primarily longitudinal and primarily transverse modes, as proposed in \cite{PhysRevB.72.205433}. A brief numerical section demonstrates the most important results of our theoretical efforts, where we use parameters for a 100 nm Ag sphere. In this section we also discuss the frequency shifts for spheroidal modes with the interesting result, that for certain frequencies the viscosity leads to an increase of the frequency compared with the ones without damping. This is difficult to explain, in particular because not all frequencies are affected in a similar manner, but we believe that the fact that spheroidal modes are neither pure longitudinal nor pure transverse vibration modes, could  play a role. This finding obviously is also interesting from a purely theoretical point of view, and further studies in this direction are necessary, also in lower dimensions (2--d and 1--d cases). 
Since the damping rates obtained by different means correspond to the lifetime of the respective excitations, the presented formulas could have an impact on experimental studies. It is furthermore possible to straightforwardly incorporate also energy loss via interaction of the sphere with its surrounding by means of additional of damping rates.   
\section*{Acknowledgment}
Markus Wenin would like to dedicate this work to the memory of his dear friend Christian Peintner (October 1972 -- December 2023). 
\appendix

\section{Outline of the solution process for the vibration modes}\label{appa}
When the complex wave numbers Eq.~\eqref{utomega}, Eq.~\eqref{ulomega} and elastic moduli, Eq.~\eqref{complexlambda} are introduced, the solution process is similar to the one without dissipation. One introduces the Helmholtz potentials, and then derives the displacement fields \cite{Eringen,Ticam}. We use the following expressions, where the prefactors $R^2$, $R$, $R^2$ have been introduced to render the constants $A_l$, $B_t$, $C_t$ dimensionless, 
\begin{equation}
\label{p1}
\Phi = A_l R^2 j_n(k_l r) P^m_n(\cos\theta)e^{i m \phi}~,
\end{equation}
\begin{equation}
\label{p2}
\Psi = B_t R j_n(k_t r)P^m_n(\cos\theta)e^{im\phi}~, 
\end{equation}
\begin{equation}
\label{p3}
\chi = C_t R^2 j_n(k_t r)P^m_n(\cos\theta)e^{i m\phi}~.
\end{equation}
The latter are necessary to derive the vectorial eigenfunctions. Additionally, we introduce the vector potential, constructed by means of $\Psi$ and $\chi$,
\begin{equation}
\label{pv}
\mathbf{\Phi}_V = r\Psi \mathbf{e}_r + \nabla\times(r\chi \mathbf{e}_r)~.
\end{equation}
With the help of these potentials, the longitudinal and transverse displacements are given by
\begin{equation}
\mathbf{u}_l(r,\theta,\phi) = \nabla \Phi~,~\mathbf{u}_t(r,\theta,\phi) = \nabla \times \mathbf{\Phi}_V~,
\end{equation}
and the complete field, Eq.~\eqref{udef}, by the sum
\begin{equation}
\mathbf{u} = \mathbf{u}_l+\mathbf{u}_t~.
\end{equation}
We construct the complete stress tensor $\sigma$, evaluate the boundary condition, and formulate a homogeneous linear system of equations with coefficient matrix $M$ by extraction of the coefficients 
$A_l$, $B_t$, $C_t$. The investigation of the determinant of $M$ shows, that $\det(M)$ can be factorized as follows (see in \textit{DampingViscoelasticSphere.nb}). A factor $e^{i m \phi}\neq 0$, a function  $f_0(\theta)\neq 0$, and two further functions $f_{1}(\tilde{\omega})$, $f_{2}(\tilde{\omega})$. One of the functions leads to the torsional modes, the other to the spheroidal modes. Because our result for spheroidal modes is not valid for $n=0$ (breathing modes), we consider this case by direct inspection of the stress tensor and the boundary condition. 
The intermediate results and proofs are extensive and not suitable to be report here, for details see our notebook \textit{DampingViscoelasticSphere.nb}.\\
An investigation of the displacement fields, Eq.~\eqref{ubreath},  Eqs.~(\ref{urspher}--\ref{uphspher}) and Eqs.~(\ref{urtors}--\ref{uphtors}) shows, that they possess the symmetry (using the magnetic quantum number $m$ as an argument to simplify the notation)
\begin{equation}
\mathbf{u}(-m)=(-1)^m \frac{(n-m)!}{(n+m)!}\mathbf{u}^*(m)~.
\end{equation}
This follows from using the relation for the associated Legendre polynomials of negative order, $P^{-m}_n(x)=(-1)^m\frac{(n-m)!}{(n+m)!} P^m_n(x)$, $m\geq 0$, and the relation for the derivative \cite{Abramowitz},
\begin{equation}\label{dPdth}
 \csc \theta  \big[(n+1) \cos \theta P_n^m(\cos \theta)+(m-n-1) P_{n+1}^m(\cos \theta)\big] \equiv -\frac{d P_n^m(\cos \theta)}{d\theta}~,
\end{equation}
when applied to Eq.~\eqref{uthspher} and Eq.~\eqref{uphtors}. Furthermore, a look at Eq.~\eqref{gV} and Eq.~\eqref{gT} shows the relations 
\begin{equation}
\gamma_{\mathrm{diss,visc}}(-m)=\gamma_{\mathrm{diss,visc}}(m)~,~\gamma_{\mathrm{diss,T}}(-m)=\gamma_{\mathrm{diss,T}}(m)~.
\end{equation}
We were not able to show the complete independence of $m$, but we believe this is of minor interest. 
\section{Auxiliary functions to determine $\gamma_{\mathrm{diss,visc}}$ for spheroidal modes}\label{appb}
The dimensionless auxiliary functions are defined by
\begin{equation}
\psi_L(n,\omega) = \frac{\psi_1(n,\omega)}{D(n,\omega)}~,~\psi_T(n,\omega)=\frac{\psi_2(n,\omega)}{D(n,\omega)}~,
\end{equation}
where the following expressions are used (with $\nu:=c_t/c_l$)

\begin{eqnarray}
\psi_1(n,\omega) = \nu ^3 \Bigg\{4 \nu  \left(n^2+n-2\right) R_t^2 J_{n+\frac{5}{2}}\left(R_l\right) J_{n+\frac{3}{2}}\left(R_t\right)- 4 \nu  R_t J_{n+\frac{1}{2}}\left(R_l\right) \times\nonumber{}\\
\Bigg[\left(n^2+n-2\right) R_t
   J_{n+\frac{3}{2}}\left(R_t\right)+\Big(2-(n+2) n^2+n+R_t^2\Big) J_{n+\frac{1}{2}}\left(R_t\right)\Bigg]+\nonumber{}\\
J_{n+\frac{1}{2}}\left(R_t\right) \Bigg[4 \nu  R_t \Big(2-(n+2) n^2+n+R_t^2\Big)
   J_{n+\frac{5}{2}}\left(R_l\right)+\nonumber{}\\
\Big(2 (2n^2-n-5) R_t^2+8 (n-1) (n+1) (n+2)-R_t^4\Big) J_{n+\frac{3}{2}}\left(R_l\right)\Bigg]+\nonumber{}\\
2 R_t \Big(8-2 (n+3) n^2+R_t^2\Big) J_{n+\frac{3}{2}}\left(R_l\right)
   J_{n+\frac{3}{2}}\left(R_t\right)+\nonumber{}\\
R_t J_{n-\frac{1}{2}}\left(R_l\right) \Bigg[R_t \left(2-4 n^2+2 n+R_t^2\right) J_{n+\frac{1}{2}}\left(R_t\right)-2 \left(R_t^2-2 n \left(n^2+n-2\right)\right)
   J_{n+\frac{3}{2}}\left(R_t\right)\Bigg]\Bigg\}~,
\end{eqnarray}

\begin{eqnarray}
\psi_2(n,\omega)= R_t J_{n-\frac{1}{2}}\left(R_t\right) \Bigg[R_t \left(2-4 n^2+2
   n+R_t^2\right) J_{n+\frac{1}{2}}\left(R_l\right)-4 \nu 
   \Big(2-(n+2) n^2+n+R_t^2\Big)
   J_{n+\frac{3}{2}}\left(R_l\right)\Bigg]+\nonumber{}\\
J_{n+\frac{1}{2}}\left(R_l\right) \Bigg[\Big(4 (n-1) \left(2 n^3+n^2-6n-6\right)+2
   (n+2) (2 n+1) R_t^2-R_t^4\Big) J_{n+\frac{3}{2}}\left(R_t\right)-\nonumber{}\\
4 R_t \Big(2 (n+1) (n-1)^2+R_t^2\Big)
   J_{n+\frac{5}{2}}\left(R_t\right)\Bigg]+\nonumber{}\\
4 \nu  R_t
   J_{n+\frac{3}{2}}\left(R_l\right) \Bigg[2 n (n+1) R_t
   J_{n+\frac{5}{2}}\left(R_t\right)+\Big(R_t^2- n (3 n^2+9n+8)\Big)
   J_{n+\frac{3}{2}}\left(R_t\right)\Bigg]~,
\end{eqnarray}
and the denominator
\begin{eqnarray}
D(n,\omega)=\nu ^2 \Bigg\{-4 \nu ^2 R_t   \left(J_{n+\frac{1}{2}}\left(R_l\right)-J_{n+\frac{5}{2}}\left(R_l\right)\right)\Bigg[\left(n^2+n-2\right) R_t
   J_{n+\frac{3}{2}}\left(R_t\right)+\nonumber{}\\
\Big(2-(n+2)n^2+n+R_t^2\Big) J_{n+\frac{1}{2}}\left(R_t\right)\Bigg]+
8 \nu  J_{n+\frac{3}{2}}\left(R_l\right) \times\nonumber{}\\
\Bigg[\Big(n (n^2+2n-1)-R_t^2-2\Big)
   J_{n+\frac{1}{2}}\left(R_t\right)-\left(n^2+n-2\right) R_t
   J_{n+\frac{3}{2}}\left(R_t\right)\Bigg]+  \nonumber{}\\
\nu  R_t
   \left(J_{n-\frac{1}{2}}\left(R_l\right)-J_{n+\frac{3}{2}}\left(R_l\right)\right) \Bigg[2\Big(2 n \left(n^2+n-2\right)-R_t^2\Big)
   J_{n+\frac{3}{2}}\left(R_t\right)+R_t \left(2-4 n^2+2 n+R_t^2\right)
   J_{n+\frac{1}{2}}\left(R_t\right)\Bigg]+\nonumber{}\\
J_{n+\frac{1}{2}}\left(R_l\right) \Bigg[R_t \Big(2 \left(4 n^2+n-5\right)-(2 n+5) R_t^2\Big)
   J_{n+\frac{5}{2}}\left(R_t\right)+\nonumber{}\\
\Big((2 n+1) (6 n+5) R_t^2-2 (n-1)
   (2 n+5) (2 n^2+4n+3)-2 R_t^4\Big)
   J_{n+\frac{3}{2}}\left(R_t\right)\Bigg]+\nonumber{}\\
4 \nu  R_t
   J_{n+\frac{3}{2}}\left(R_l\right) \Bigg[2 n (n+1) R_t
   J_{n+\frac{5}{2}}\left(R_t\right)+\Big(n (n^2+2n-1)-R_t^2-2\Big)
   J_{n-\frac{1}{2}}\left(R_t\right)+\nonumber{}\\
\Big(R_t^2-n (3 n^2+9n+8)\Big)
   J_{n+\frac{3}{2}}\left(R_t\right)\Bigg]\Bigg\}~.
\end{eqnarray}

\bibliography{referenzliste}

\end{document}